# C#: Connecting a Mobile Application to Oracle Server via Web Services

**Daniela Ilea,**
Senior software developer
Microsoft Certified Professional

ABSTRACT: This article is focused on mobile development using Visual Studio 2005, web services and their connection to Oracle Server, willing to help programmers to realize simple and useful mobile applications.

## 1. Introduction

As a user and a programmer of mobile application, I don't like to be limited to Microsoft products, but be able to use Oracle database, for example.

Since the appearance of Oracle XE, many developers took in consideration this option, because of its advantages:
- Oracle XE is free to develop, deploy, and distribute;
- XE will store up to 4GB of user data;
- Oracle Database XE can be installed on any size host machine with any number of CPUs (one database per machine).

Oracle does not offer a driver for accessing Oracle server through Visual Studio 2005 mobile applications, and the solutions found on the internet are not free. So, one solution would be using Web Services.

Web services provide a way of linking through a network different modules developed different platforms using different languages.

Regarding web services:
- The communication between server and client application is based on a standard format called XML (eXtensible Markup Language) universal accepted through platforms.

101

- WSDL (Web Service Description Language) is a file located on application folder containing information about namespace of the XML file and the description of the elements that the service consists in.
- SOAP (Simple Object Access Protocol) is the protocol of communication with the web server.
- IIS (Internet Information Server), the web server.

Our application will need the following:
- Visual Studio 2005
- IIS
- Oracle XE Server

The applications will be structured as in the following figure:

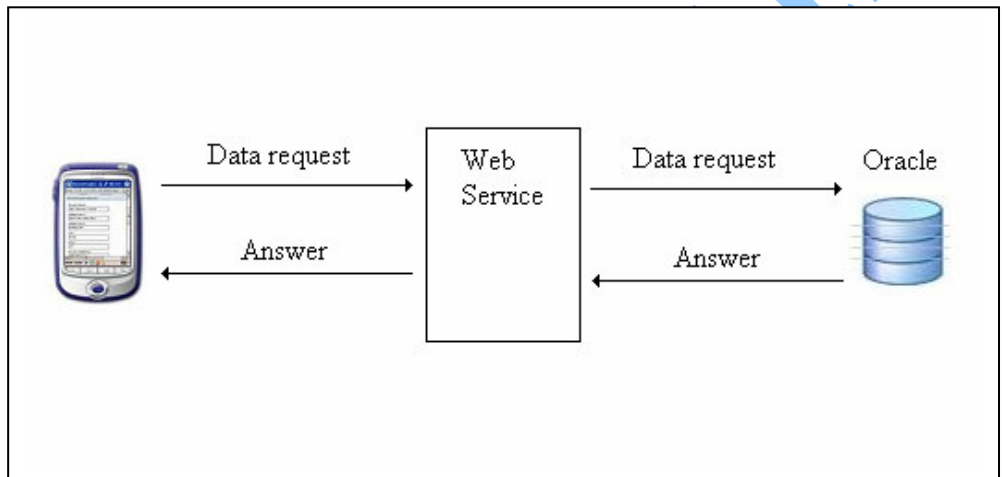

## 2. Oracle Server

After the server is installed, we need to create the Oracle objects for our application:
- tables to hold our data
- Packages with procedures for managing our data

Of course as the complexity of application is increased, the structure of the database is more complex. But for our sample purpose, those two will be enough.

## 3. IIS

The IIS needs some configuration in order to host our web service.
- For the web service a virtual folder is created. In this folder the web service will be deployed.
- Set the rights for the virtual folder (read, execute..)
- Other web server specific configuration

## 4. Web Service

The web service is created as a distinct website application using Visual Studio .NET 2005.

First of all, the web service needs to have a reference to *"Oracle.DataAccess.dll"*

Through the AppCode folder , I created a separate class containing methods necessary to communicate with Oracle.

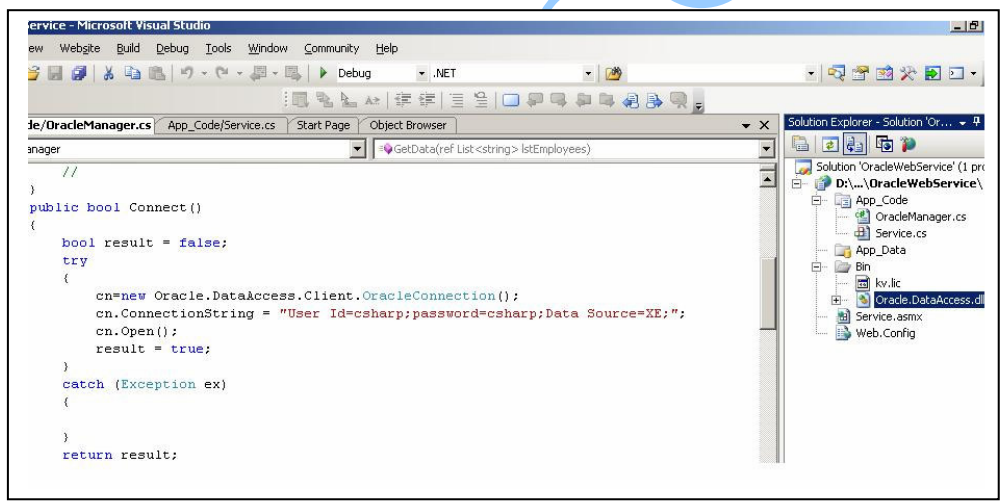

We need a connection defined as following:

*Oracle.DataAccess.Client.OracleConnection cn;*
*cn=new Oracle.DataAccess.Client.OracleConnection();*
*cn.ConnectionString = "User Id=csharp;password=csharp;Data Source=XE;";*
*cn.Open();*

103

To hold the data gathered from Oracle tables, I used a generic list (a strongly typed list of objects that can be accessed by index).

I read the data from Oracle by using an *OracleCommand* and *OracleDataReader*.

The WebServices methods are exposed to the client application through Service.cs class.

Each method in a web service needs to have the tag:

*[WebMethod]*
*public void GetEmployeesData(ref List<string> strData)*
 *{*
   *manager = new OracleManager();*
   *manager.GetData(ref strData);*
 *}*

After the web service is created and compiled, it needs to be published onto the webserver (IIS) in the virtual folder previously created.

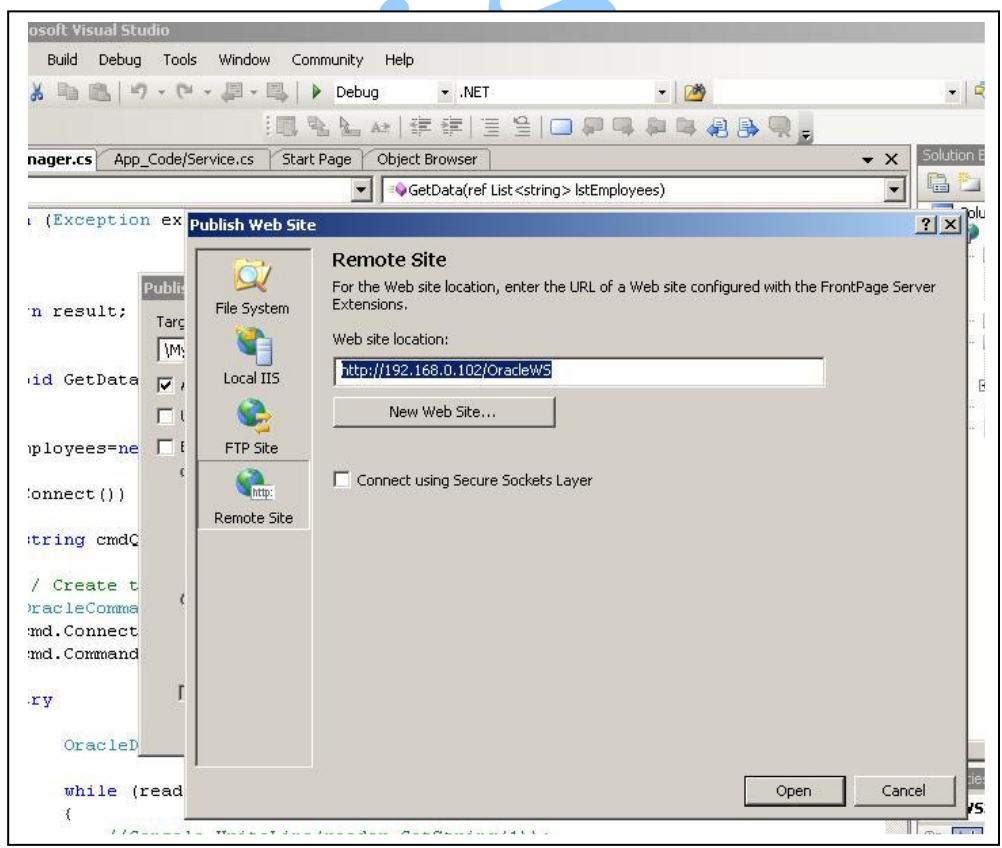

Testing the web service can be done by using a browser, typing the web address followed by the Service application.
Exemple:   http://localhost:1576/OracleWebService/Service.asmx,   where OracleWebService is the virtual folder and Service is the web service.
After that step, the web service is ready to be consumed(is ready to accept requests from the client application).

## 5. Mobile application

The mobile application needs to have an web reference to the web service previously created.
After declaring an object of type *OracleWS.Service* all the methods from the web service are available.

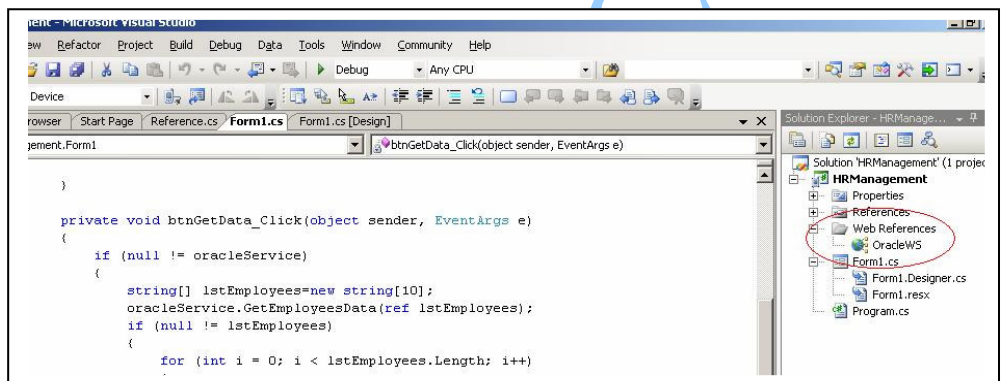

## Conclusions

Web services are easy to use and implement and they represent my choice as a Visual Studio programmer
There are a few other ways (some may say better) :
- We can use PHP web server and PHP for an web service
- We can use Apache Tomcat and java server pages for web service